\documentclass{WileyMSP-template}
\usepackage{amsmath}

\begin{document}

\pagestyle{fancy}

\title{Pseudo-waveform-selective metasurfaces and their limited performance}

\maketitle


\author{Tomoyuki Nakasha}
\author{Sendy Phang}
\author{Hiroki Wakatsuchi*}


\dedication{}

\begin{affiliations}
T. Nakasha, Prof. H. Wakatsuchi\\
Department of Electrical and Mechanical Engineering, Graduate School of Engineering, Nagoya Institute of Technology, Aichi, 466-8555, Japan\\
wakatsuchi.hiroki@nitech.ac.jp:

Prof. S. Phang\\
George Green Institute of Electromagnetic Research, Faculty of Engineering, The University of Nottingham, Nottingham NG7 2RD, UK

Prof. H. Wakatsuchi\\
Precursory Research for Embryonic Science and Technology (PRESTO), Japan Science and Technology Agency (JST), Saitama 332-0012, Japan

\end{affiliations}


\keywords{electromagnetics, microwaves, metamaterials, metasurfaces, waveform, pulse width}

\begin{abstract}

In recent years, metasurfaces composed of lumped circuit components, including nonlinear Schottky diodes, have been reported to be capable of sensing particular electromagnetic waves even at the same frequency depending on their waveforms, or more specifically, their pulse widths. In this study, we report analogous waveform-selective phenomena using only linear circuits and linear media. Although such linear metasurfaces are analytically and numerically demonstrated to exhibit variable absorption performance, it cannot strictly be categorized as waveform-selective absorption. It is due to the fact that the waveform-selective responses in the linear metasurfaces are originated from the dispersion behaviors of the structures rather than the frequency-conversion seen in nonlinear waveform-selective metasurfaces. We thus refer to these linear structures as pseudo-waveform-selective metasurfaces. Additionally, we show that the pseudo-waveform-selective metasurfaces have limited performance unless nonlinearity is introduced. These results and findings confirm the advantages of nonlinear waveform-selective metasurfaces, which can be exploited to provide an additional degree of freedom to address existing electromagnetic problems/challenges involving even waves at the same frequency.

\end{abstract}


\section{Introduction}
Artificially engineered subwavelength periodic structures, or so-called metamaterials \cite{smithDNG1D, smithDNG2D2} and metasurfaces \cite{EBGdevelopment}, are well-known to exhibit a wide range of electromagnetic properties, including negative permittivity \cite{pendryENG1, pendryENG2}, negative permeability \cite{pendryMNG}, negative refractive indices \cite{smithDNG2D2}, zero refractive indices \cite{ziolkowski2004propagation}, asymmetrical responses \cite{phang2017theorya}, and extremely large surface impedances \cite{EBGdevelopment}. These artificially tailored properties have been exploited to design advanced electromagnetic devices and applications such as perfect lenses \cite{pendryperfetLenses, nzSuperlens, fangSuperlens, germanHyperLens}, cloaking devices \cite{enghetaCloaking, pendryCloaking, mantleCloaking, cloakingSensor}, wavefront shaping devices \cite{yu2011light, pfeiffer2013metamaterial, yu2014flat}, ultrathin absorbers \cite{enghetaAbs, ultraThinAbs, mtmAbsPRLpadilla, My1stAbsPaper, huang2019catenary, watts2012metamaterial}, and antennas \cite{sievTunableImpedance, ziolkowski2011metamaterial}. These capabilities can be further extended or improved by introducing nonlinearity into metamaterials and metasurfaces \cite{kivshar2003nonlinear, yang2015nonlinear, cai2018multifunctional, wang2020saturable}. For example, nonlinear metasurface absorbers \cite{aplNonlinearMetasurface, kim2016switchable, li2017high, li2017nonlinear} permit the transmission of low-amplitude signals used for antenna communications while effectively absorbing strong electromagnetic noise that may affect sensitive electronics even at the same frequency, as opposed to linear metasurface absorbers, which accept both types of waves. Recently, circuit-based metasurfaces containing nonlinear Schottky diodes have been reported to discriminate particular electromagnetic waves at the same frequency depending on their waveforms, or more specifically, their pulse widths \cite{wakatsuchi2013waveform, eleftheriades2014electronics, wakatsuchi2015waveformSciRep, wakatsuchi2019waveform, vellucci2019waveform}. Such waveform-selective metasurfaces are expected to provide an additional degree of freedom to address a wide range of existing problems/challenges involving waves at the same frequency, for example, in electromagnetic interference \cite{wakatsuchi2019waveform}, antenna design \cite{vellucci2019waveform}, and wireless communications \cite{wakatsuchi2015waveformSciRep, ushikoshi2019experimental}. Although all waveform-selective metasurfaces reported to-date require the use of an electromagnetic nonlinear effect, here we demonstrate that similar phenomena can be obtained using only linear circuits and linear media. This is because the frequency content of a pulsed sinusoidal wave depends on its pulse width. In this study, we report such simple linear metasurfaces that exhibit varying absorption performance depending on the pulse width of the incoming electromagnetic wave. Importantly, we both analytically and numerically show that the observed absorption behaviors of the linear metasurfaces are based on a totally different mechanism, compared to that of the nonlinear waveform-selective metasurfaces; thereby we refer to this kind of absorption as \emph{pseudo}-waveform-selective absorption. In addition, these pseudo-waveform-selective metasurfaces are shown to have limited performance compared to nonlinear structures. 

\section{Nonlinear waveform-selective metasurfaces}

The conventional nonlinear waveform-selective absorption mechanism is achieved in metasurfaces by using both rectification to zero frequency and transient responses in the time domain \cite{wakatsuchi2015waveformSciRep, wakatsuchi2019waveform}. The former frequency conversion is achieved by deploying diodes across gaps between conductor edges. In particular, if four diodes form a diode bridge, more of the incoming wave energy is converted to zero frequency than in the case of half-wave rectification \cite{aplNonlinearMetasurface, wakatsuchi2013waveform}. In addition, if an inductor is connected to a resistor in series within the diode bridge, a strong electromotive force appears, which prevents incoming electric charges during its initial time period. However, this force is gradually reduced due to the zero-frequency component of the electric charges. Consequently, a short pulse is not effectively absorbed, while the energy of a long pulse or a continuous wave (CW), even at the same frequency, is strongly dissipated by the series resistor. 

\begin{figure}
\includegraphics[width=0.48\linewidth]{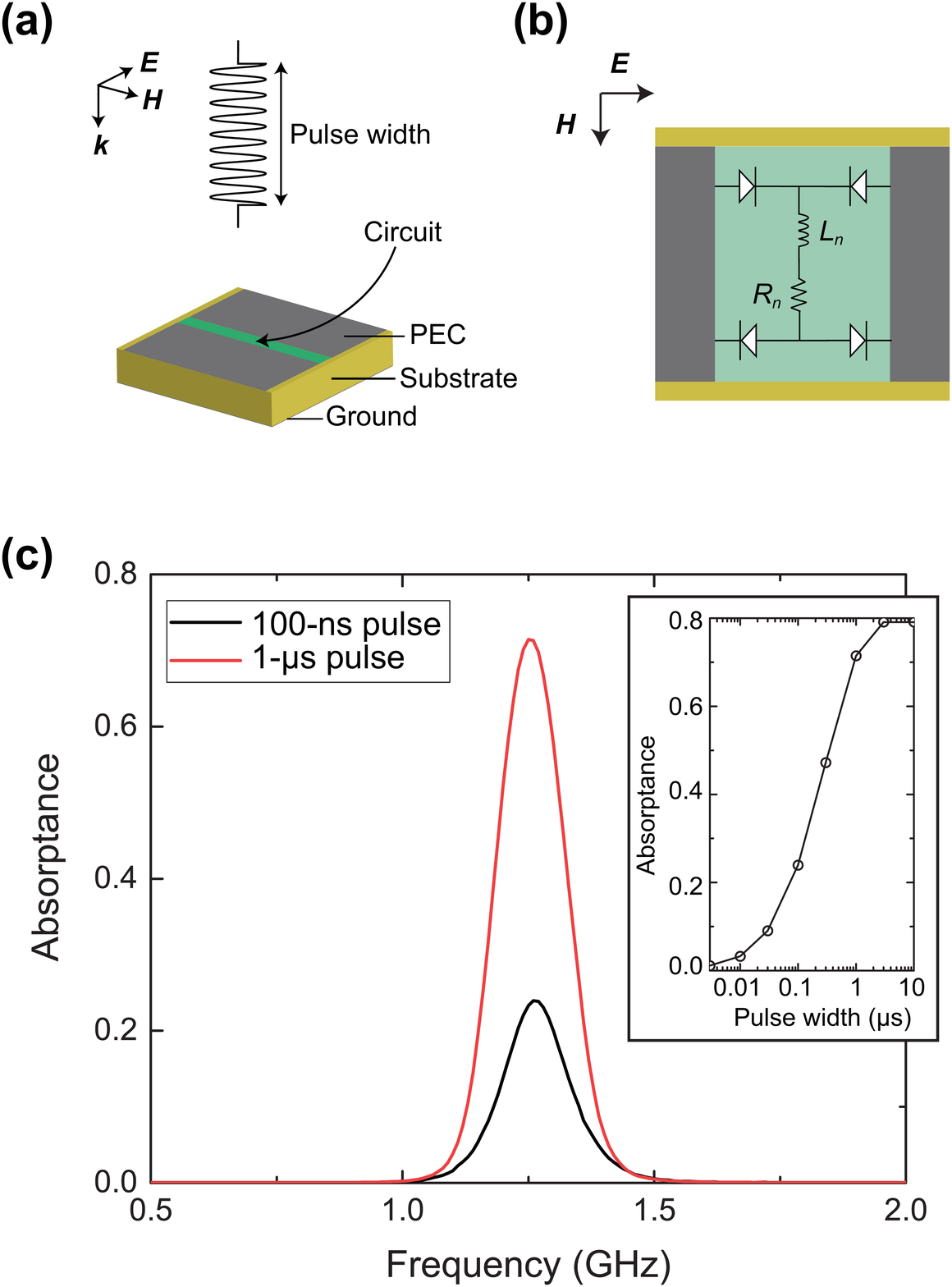}
\caption{(a) A periodic unit cell receiving an incident pulsed wave. (b) Circuit configuration for a nonlinear waveform-selective metasurface ($L_n$ = 100 ${\rm \mu}$H and $R_n$ = 10 $\Omega$). (c) Absorption performance for pulsed waves. The inset shows the pulse width dependence of the structure at 1.25 GHz. The input power level was set to 0 dBm. }
\label{fig.1}
\end{figure}

\begin{figure}
\includegraphics[width=0.48\linewidth]{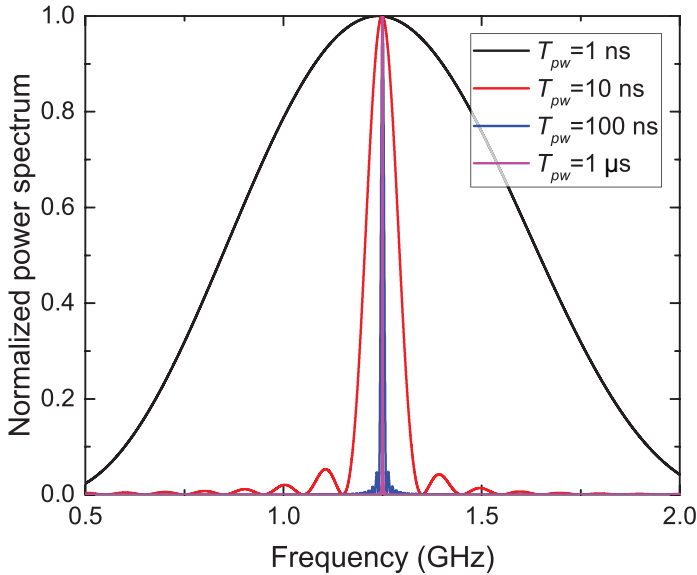}
\caption{Spectra of pulsed cosine waves with different pulse widths. The oscillation frequency was set to 1.25 GHz. }
\label{fig.2}
\end{figure}

This is numerically demonstrated in Figure \ref{fig.1}, where the depicted waveform-selective metasurface was simulated using a co-simulation method based on ANSYS Electronic Desktop (R18.1) \cite{wakatsuchi2013waveform}. Note that since eventually this simulation method produces results on a circuit simulator, transient electromagnetic field distributions cannot be visualized unlike ordinary simulations. Also, this simulation modeled diodes based on a SPICE model provided by Broadcom (specifically, the HSMS-286x series but without series resistance or junction capacitance for simplicity). The conducting patches were designed to be 46 mm $\times$ 46 mm with 47 mm periodicity. The substrate was set to be 5.6 mm thick (Rogers 3003 without dielectric loss) on a ground plane. As seen from Figure \ref{fig.1}c, the nonlinear waveform-selective metasurface absorbed a 1-$\mu$s pulse more strongly than a 100-ns pulse at the same frequency of 1.25 GHz. The inset of Figure \ref{fig.1}c shows the absorption performance with respect to the pulse width. It is important to note that the spectrum of the 1-$\mu$s pulse and that of the 100-ns pulse are almost the same, as shown in Figure \ref{fig.2}, where each power spectrum $P_{PLS}$ was obtained using the following equation \cite{wakatsuchi2015time}:
\begin{eqnarray}
\label{eq.pulseSpectrum}
E_{PLS}(f)= \displaystyle \frac{a}{2}\Biggl( \displaystyle \frac{\sin{(2\pi (f -f_0)T_{pw})}}{\pi (f -f_0)}+\displaystyle \frac{\sin{(2\pi (f +f_0)T_{pw})}}{\pi (f +f_0)}\Biggr).
\end{eqnarray}
Equation (\ref{eq.pulseSpectrum}) assumes that the incident pulse is based on a cosine function. In addition, $E_{PLS}$, $a$, $f_0$, and $T_{pw}$ represent the electric field, magnitude, oscillation frequency, and pulse width of the incident pulse, respectively. Other variables, including a phase delay and the location of the point at which the wave is observed, are omitted for simplicity. Despite how small the difference is between the spectra of $T_{pw}$ = 100 ns and 1 $\mu$s, the nonlinear waveform-selective metasurface was capable of varying absorptance. 

Another important aspect in Figure \ref{fig.2} is that the spectrum of a shorter pulse appears totally different. For instance, the 1-ns pulse contains nonnegligible frequency components in addition to the oscillation frequency due to the discontinuity of the waveform at its beginning and end. Such spectral differences are exploited below.  

\section{Linear pseudo-waveform-selective metasurfaces}
\begin{figure}
\includegraphics[width=0.48\linewidth]{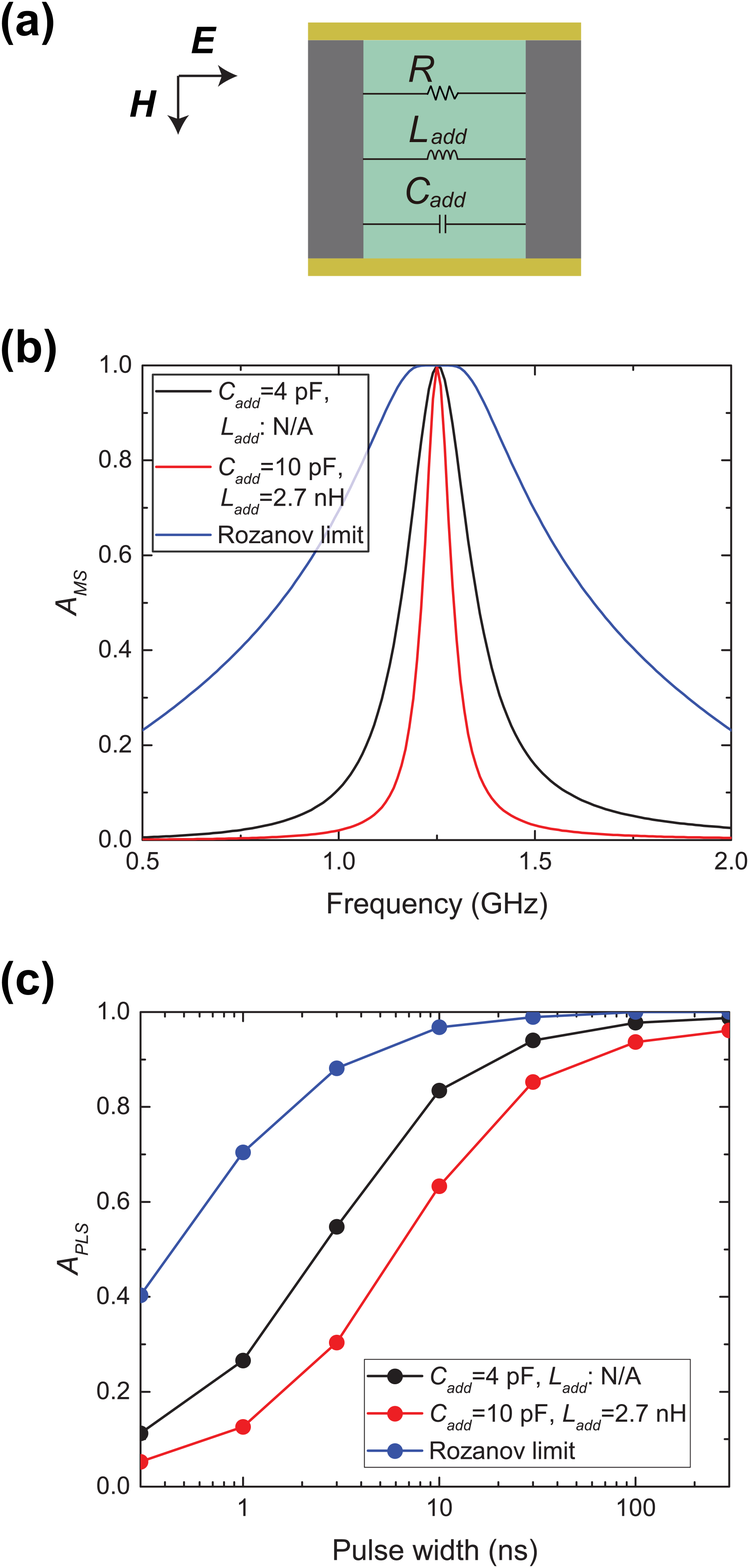}
\caption{(a) Circuit configuration used for a linear pseudo-waveform-selective metasurface. $R$ is fixed at $\approx$ 377 $\Omega$. (b) Absorption performance for a continuous sine wave. (c) Absorbed energy of pulsed sine waves with $f_0$ set to 1.25 GHz. }
\label{fig.3}
\end{figure}

As opposed to the above structure, which contains nonlinearity, the metasurface reported in this study is composed of ideally linear circuits and linear media, although in practice, nonlinearity may need to be included to some extent. More specifically, its periodic unit cell consists of a ground plane (perfect electric conductor: PEC), a dielectric substrate (3-mm-thick Rogers 3003 but without dielectric loss for simplicity), conducting patches (PEC, 9 mm $\times$ 9 mm), and a lumped capacitor, inductor, and resistor with 10 mm periodicity (see Figure \ref{fig.3}a). Similarly to other general metasurfaces \cite{ultraThinAbs, mtmAbsPRLpadilla, My1stAbsPaper}, this type of structure resonates at a designed frequency and strongly absorbs an incoming frequency within its bandwidth $BW$, which depends on the total effective capacitive component $C$ and the total effective inductive component $L$, i.e., 
\begin{equation}
\label{eq:BW}
BW \propto \sqrt{L/C}. 
\end{equation}
$C$ and $L$ can be adjusted not only by means of a periodically metallized pattern containing geometric capacitive and inductive components $C_0$ and $L_0$ but also by lumped circuit elements \cite{baena2005equivalent, ZhouCWeq, MyCWeqCircuitPaper}. For instance, Figure \ref{fig.3}a shows a periodic unit cell of the linear structure tested using a lumped capacitor $C_1$ and inductor $L_1$ that are deployed between two adjacent patches. As seen in Figure \ref{fig.3}b, where ordinary small-signal analysis (assuming a sine wave as the incident waveform) is performed, increasing the value of $C$ produces narrower $BW$. This is because this structure can be effectively represented by a parallel circuit and $C=C_0+C_1$, while $L^{-1}=L_0^{-1}+L_1^{-1}$ (see also Equation (\ref{eq:BW})). Related to its intrinsic absorption performance, the structure exhibits varying absorptance for pulsed sine waves with different pulse widths. As plotted in Figure \ref{fig.3}c, where the input waveform is changed to a pulsed sine wave ($f_0$ = 1.25 GHz) with different pulse widths, the absorptance of the structure appears to reach 1.0 more slowly when a larger value of capacitance is used (compare the black curve to the red curve). Note that in both cases, the resonant frequencies are set to almost the same value (i.e., 1.25 GHz) to ensure a fair comparison. More importantly, this is not a waveform-selective absorption mechanism, which is defined as an absorption mechanism that varies in response to the waveform or pulse width of an incoming wave, even at a `\emph{constant}' frequency. Instead, in the above case, the structures see the difference in the frequency spectrum (compare the pulse width range in Figure \ref{fig.3}c to the spectra of the corresponding pulse widths in Figure \ref{fig.2}). For this reason, the vertical axis of Figure \ref{fig.3}c is changed to $A_{PLS}$, which is distinct from $A_{MS}$, which is the absorptance of the metasurface itself. Although explained in detail below, $A_{MS}$ represents absorptance at a `\emph{constant}' frequency as commonly calculated by $A_{MS}=1-|S_{11}|^2$, while $A_{PLS}$ is not necessarily a ratio at a constant frequency. Also, Figure \ref{fig.3}c can be compared to the inset of Figure \ref{fig.1}c, where nonlinear waveform-selective metasurfaces similarly varied absorptance but for longer pulse widths. 

To derive $A_{PLS}$, suppose that $b^+$ and $b^-$ denote an incident waveform and a reflected waveform, respectively. Then, the energy spectral density of a reflected pulse $E_{r}$ is calculated as 
\begin{eqnarray}
E_{r}=b^-(t)\cdot \left\lbrace b^-(t)\right\rbrace ^\dagger,
\end{eqnarray}
where $\dagger$ represents the conjugate transpose. Based on Parseval's theorem, 
\begin{eqnarray}
E_{r}=\int |b^-(f)|^2df=b^-(f)\cdot \left\lbrace b^-(f)\right\rbrace ^\dagger.
\end{eqnarray}
Hence, the normalized energy spectral density of the reflected pulse, $R_{PLS}$, is
\begin{eqnarray}
R_{PLS}=\frac{\int |b^-(f)|^2df}{\int |b^+(f)|^2df}=\frac{\int |b^+(f)S_{11}(f)|^2df}{\int |b^+(f)|^2df}.
\end{eqnarray}
Since the normalized energy spectral density of the corresponding absorbed pulse, $A_{PLS}$, is $A_{PLS}=1-R_{PLS}$,
\begin{eqnarray}
\label{eq.Apls}
A_{PLS}=1-\frac{\int {|b^+(f)\cdot S_{11}(f)|^2}df}{\int {|b^+(f)|^2}df}.
\end{eqnarray}
According to this equation, $A_{PLS}$ varies with $b^+$. This absorption mechanism is different from the case of waveform-selective absorption, in which it is assumed that the incoming spectrum remains at almost a constant value (or at least is sufficiently narrow compared to the bandwidth of the structure). Nonlinear waveform-selective metasurfaces primarily exhibit variation in $S_{11}$ in the time domain. In contrast, for linear metasurfaces, it is impossible for $A_{PLS}$ to vary because $S_{11}$ remains the same unless $b^+$ itself changes. For this reason, we refer to the variations seen in Figure \ref{fig.3}b as a `\emph{pseudo}'-waveform-selective absorption mechanism. 

Additionally, we note that such linear structures have limited performance. First, there is a limit to the variation of $A_{PLS}$ in a small pulse width range if the thickness $d$ of the entire structure is fixed. This performance is limited by the so-called Rozanov limit \cite{rozanov2000ultimate}, i.e., 
\begin{equation}
\label{eq.1}
\Delta \lambda<\frac{2\pi^2 d}{|\ln{S_{11}}|},
\end{equation}
where $\Delta \lambda$ represents the operating bandwidth associated with wavelength, i.e., $\Delta \lambda = \lambda_1-\lambda_2$. Under the assumptions that $A_{MS}=1-|S_{11}|^2$, $\lambda=c/f$, where $c$ is the speed of light, and the operating frequency $f_0$ is the central value between $f_1$ and $f_2$ (i.e., $f_0-f_1=f_2-f_0$), Equation (\ref{eq.1}) leads to 
\begin{equation}
\label{eq.2}
A_{MS}<1-\exp{\Biggl(\pm \frac{4\pi^2 d}{\displaystyle \frac{c}{f_{1}}-\displaystyle\frac{c}{2f_{0}-f_{1}}}\Biggr)}.
\end{equation}
Equation (\ref{eq.2}) indicates that $A_{MS}$ (i.e., the absorptance of the considered metasurface) does not have a wider bandwidth than the right side of the equation. By equating the left side of Equation (\ref{eq.2}) to its right side, the Rozanov limit is plotted in Figures \ref{fig.3}b and c. As seen in these results, no linear structure tested here exceeded the Rozanov limit. 

\begin{figure}
\includegraphics[width=0.48\linewidth]{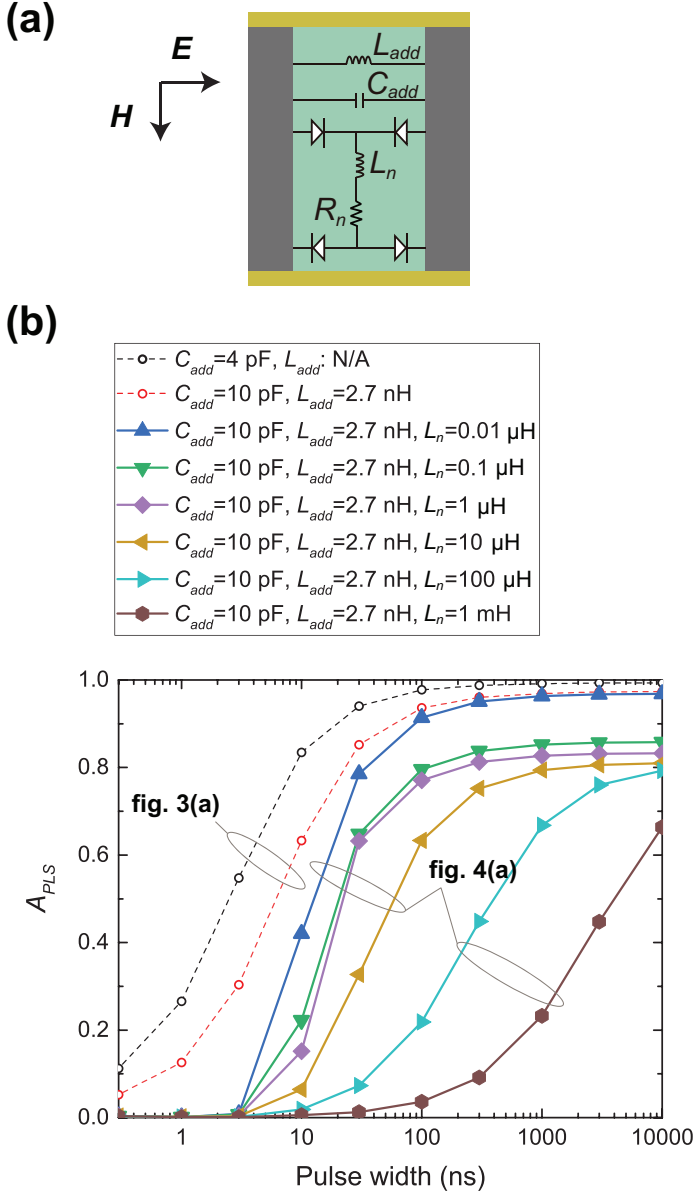}
\caption{(a) Circuit configuration used for nonlinear waveform-selective metasurfaces. (b) Absorbed energy of pulsed sine waves with $f_0$ and the input power set to 1.25 GHz and 0 dBm, respectively. $R$ and $R_n$ are fixed at $\approx$ 377 $\Omega$ and 10 $\Omega$, respectively. Note that these nonlinear waveform-selective metasurfaces have the same physical dimensions as the ones used for the linear metasurfaces in Figure \ref{fig.3}.}
\label{fig.4}
\end{figure}

In addition, linear structures also have poorer performance than nonlinear waveform-selective metasurfaces in terms of controlling long pulses. The pulse energies absorbed by the above linear structures are compared to those absorbed by nonlinear waveform-selective metasurfaces in Figure \ref{fig.4}. Both types of structures have the same physical dimensions and almost the same operating frequencies with slightly different absorption strengths. However, the nonlinear waveform-selective metasurfaces tend to show variations in absorptance over a broad pulse width range. In fact, this range depends on what circuit values are used for the internal inductor and resistor ($L_n$ and $R_n$), which determine the time constant $\tau$ of the structure in Figure \ref{fig.4}b approximately as follows \cite{aplEqCircuit4WSM}: 
\begin{equation}
\label{eqn:tau}
\tau=\frac{L_n}{R_n+R_d},
\end{equation}
where $R_d=2R_0$ and $R_0$ is the effective resistive component of one of the diodes used. Note that Equation (\ref{eqn:tau}) is based on a simplified equivalent circuit model to predict time-domain responses \cite{aplEqCircuit4WSM}. In this study, this equation is not used for our numerical simulations, where nonlinear diodes are modeled using SPICE parameters. However, this equation can be used as a design guide to see how time constant relates to $L_n$, $R_n$, and $R_d$ ($R_d$ is obtained by different methods including a current-voltage curve of the specific diode used \cite{aplEqCircuit4WSM}). For instance, the time constant of a nonlinear waveform-selective metasurface varies with increasing $L_n$. In particular, the time constant of the nonlinear waveform-selective metasurface with $L_n$ = 10 $\mu$H to 1 mH is proportionally scaled from 100 ns to 10 $\mu$s in Figure \ref{fig.4}b. Interestingly, however, such scalability of the time constant disappears as $L_n$ decreases from 10 $\mu$H, causing the curve of the nonlinear waveform-selective metasurface to approach that of the linear metasurface with the same $L_{add}$ and $C_{add}$. This is presumably because $L_n$ and the diodes are almost short-circuited here and the nonlinear metasurface behaves like an ordinary linear metasurface. Nonetheless, Figure \ref{fig.4}b demonstrates that nonlinear waveform-selective metasurfaces are readily capable of exhibiting absorptance variations even over a broad pulse range with the replacement of lumped inductor $L_n$.  

\section{Discussion}
The Rozanov limit mentioned in Equations (\ref{eq.1}) and (\ref{eq.2}) and Figure \ref{fig.3} is known to be broken with the introduction of nonlinearity. For instance, the use of non-Foster circuits, which do not conform to Foster's reactance theorem \cite{linvill1953transistor, aberle2007antennas}, allows the intrinsic capacitive and inductive components of a metasurface, which are strongly frequency dispersive, to be canceled out. As a result, the frequency dependence of a metasurface can be mitigated to achieve a broadband response \cite{mou2016design, fan2016active}. Thus, metasurfaces loaded with non-Foster circuits can potentially control shorter pulses than the linear metasurfaces in Figure \ref{fig.3}. 

More importantly, however, even the longest pulses that can be controlled by linear metasurfaces may be too short as practical wireless communication signals. For instance, the $A_{PLS}$ of our linear structures varied in the vicinity of 10 ns, but most Wi-Fi signals generated from commercial Wi-Fi modules have pulse widths of 50 $\mu$s or longer, indicating that the applicability of linear metasurfaces in the field of wireless communications is severely limited (see \cite{ushikoshi2019experimental} for nonlinear waveform-selective metasurfaces controlling 50-$\mu$s pulses). Also, linear metasurfaces or metamaterials are known to be integrated with sensors or detectors to achieve more advanced functionalities or properties \cite{cloakingSensor}. Potentially, our linear structures can be applied to this field as well. However, a switching capability usually requires a nonlinearity and thus cannot be achieved by our linear structures unless this is a short time period as seen in Figure \ref{fig.3}c. Even in this case, broadband response is needed and cannot be readily achieved by ordinary devices such as antennas. We also note that our linear metasurfaces can be replaced by more complicated structures, but our metasurfaces can be effectively represented by a simple parallel circuit composed of a resistor, an inductor, and a capacitor, which enables us to choose a wide range of value for wave impedance, $BW$, and $Q$ factor. Another issue with linear metasurfaces is that while their $A_{PLS}$ can be easily increased, it is relatively difficult to gradually decrease $A_{PLS}$. In contrast, nonlinear waveform-selective metasurfaces offer more freedom in the design of their pulse width dependence (not only simple monotonic increases and decreases but also more complicated patterns \cite{wakatsuchi2015time}) through modification of the circuit configurations in the diode bridges. Therefore, although linear pseudo-waveform-selective metasurfaces have an advantage that ideally they are independent of the incoming power level, nonlinear waveform-selective metasurfaces have important advantages over linear metasurfaces.  

It is not impossible for nonlinear waveform-selective metasurfaces to achieve nearly 100 \% absorption for a short pulse \cite{wakatsuchi2015waveformJAP, wakatsuchi2016responses}. However, this also increases the absorptance for another type of waveform (i.e., a long pulse or CW) at the same time. Potentially, the gap between the two waveforms can be more increased by modifying the design of waveform-selective metasurface. To the best of our knowledge, however, so far none of studies has reported waveform-selective metasurfaces varying scattering parameters between nearly 0.0 and 1.0, presumably because this relates to some other loss mechanisms within substrate and circuit components (e.g., reduction in a parasitic resistance of diode contributes to increasing the difference between two waveforms as reported in the literature \cite{wakatsuchi2016responses}). Finally, the relationship between the turn-on and breakdown voltages of the diodes used is important, since this determines the dynamic range for applications of nonlinear waveform-selective metasurfaces. In general, it is difficult to adjust the power level of a received signal sent from a distant antenna in wireless communications. However, the received power level can be more easily adjusted, if the metasurfaces are deployed near a transmitter or a radiation source.  

\section{Conclusion}
We have reported the absorption performance of linear metasurfaces using pulsed incident waves. We have shown that although they behave like recently reported nonlinear waveform-selective metasurfaces, which can exhibit absorptance variations even at the same frequency, these linear structures exhibit only pseudo-waveform-selective absorption behavior. Moreover, such pseudo-waveform-selective metasurfaces are found to have limited performance unless nonlinearity is introduced. This study has clarified the performance of linear waveform-selective metasurfaces and confirmed the advantages of nonlinear waveform-selective metasurfaces, which can be exploited to provide an additional degree of freedom to address a range of electromagnetic problems/challenges involving even waves at the same frequency.

\medskip
\textbf{Acknowledgements} \par 
This study was supported in part by the Japan Science and Technology Agency (JST) under Precursory Research for Embryonic Science and Technology (PRESTO) \#JPMJPR193A. 

\medskip
\textbf{Conflict of Interest} \par 
The authors declare no conflict of interest. 

\medskip

%

\providecommand{\noopsort}[1]{}\providecommand{\singleletter}[1]{#1}%




\end{document}